\documentclass[10pt,a4paper,aps,superscriptaddress,preprint,showpacs,showkeys]{elsarticle}
\usepackage[latin9]{inputenc}
\usepackage{amsmath}
\usepackage{amssymb}
\usepackage{graphicx}
\usepackage{esint}

\makeatletter

\usepackage{a4wide}\usepackage{subfigure}\usepackage{latexsym}% useful for coding complex math

\makeatother

\begin{document}

\title{Superstatistics and the quest of generalized ensembles equivalence
in a system with long-range interactions }

\author{Nelson A. Alves$^{1}$}

\ead{alves@ffclrp.usp.br}

\author{Rafael B. Frigori$^{2}$}

\ead{frigori@utfpr.edu.br}

\address{$^{1}$Departamento de Física, FFCLRP, \\
 Universidade de São Paulo, Avenida Bandeirantes, 3900 \\
 14040-901, Ribeirão Preto, SP, Brazil. \\
 $^{2}$ Universidade Tecnológica Federal do Paraná, \\
 Rua Cristo Rei, 19 \\
 85902-490, Toledo, PR, Brazil.}
\begin{abstract}
The so-called $\chi^{2}$-superstatistics of Beck and Cohen (BC) is
employed to investigate the infinite-range Blume-Capel model, a well-known
representative system displaying inequivalence of canonical and microcanonical
phase diagrams. While not being restricted to any of those particular
thermodynamic limits, our analytical result can smoothly recover both
canonical and microcanonical ensemble solutions as its nonextensive
parameter $q$ is properly tuned. Additionally, we compare our findings
to ones previously obtained from a generalized canonical framework
named Extended Gaussian ensemble (EGE). Finally, we show that both
EGE and BC solutions are equivalent at the thermodynamic level.
\end{abstract}
\maketitle
 05.20.Gg, 05.70.Fh, 64.60.-i

\section{Introduction}

Superstatistics inception by Beck and Cohen \cite{beck-322-2003}
was intended to provide an extension of the standard statistical mechanics
formalism into a more general one, focusing on describing out-of-equilibrium
systems, which are most likely characterized by spatio-temporal fluctuations
of an intensive parameter. Its usual formulation \cite{wilk-2000,beck-87-2001,sattin-65-2002}
employs, as a working hypothesis, the argument that fluctuations evolve
on a long-time scale, while the studied system can still be locally
decomposed in many small cells (subsystems) obeying the equilibrium
statistical mechanics characterized, for instance, by an effective
local inverse temperature $\beta$. For such systems, not only the
temperature environment is considered to be a fluctuating quantity,
with probability density $f\left(\beta\right)$, but also it may carry
a spatial modulation as a classical scalar field.

Despite of some early criticism \cite{lavenda-1}, superstatistics
has been growing \cite{beck-322-2003,beck-16-2004,abe-2007,beck-39-2009,beck-369-2011}
as a consistent framework able to provide deep physical insights for
a large variety of complex nonequilibrium stationary systems. It is
corroborated by the fact that the understanding of BC formulation
of superstatistics greatly profits from the perspective of a Bayesian
formalism, as exposed by Sattin \cite{sattin-49-2006}. For instance,
let us suppose one is interested in measurements of the energy $E$
emerging as the outcome of an experiment reproducible as many times
as desired. The modeling of this experiment depends on some parameters
and assumptions related to an underlying stochastic process. So, let
$\beta$ be the parameter that mainly accounts for the dynamics of
such experiment, whose due description then implies on a proper knowledgement
of the parameter. Within the Bayesian framework, the knowledgement
about $\beta$ is cast in terms of probabilities known as the \textit{prior}
distribution $p(\beta)$ and, the \textit{posterior} distribution
described by the conditional form $p(\beta|E)$ contains all information
about $\beta$ once the observations on $E$ were given. However,
this is clearly an unknown distribution.

In this vein, superstatistics introduces a procedure to obtain marginal
probability $p(E)$ from nonequilibrium dynamical processes once given
the prior distribution $p(\beta)$ \cite{beck-322-2003,beck-87-2001},
\begin{equation}
p(E)=\int p(E|\beta)p(\beta)d\beta.\label{pe_init}
\end{equation}
This circumvents the ``learning'' from experiment $p(\beta|E)\sim p(E|\beta)p(\beta)$
\cite{sattin-49-2006,lavenda-2} by assuming that a prior distribution
is known, $p(\beta)=f(\beta)$, where $f(\beta)$ physically accounts
for the fluctuating parameter $\beta,$ whose probability distribution
function (PDF) is introduced in an \textit{ad hoc} manner and represents
our \textit{degree-of-belief}.

The distribution $p(E|\beta)$ in the BC (type-B superstatistics \cite{beck-16-2004})
formulation assumes that the thermodynamical description is statistically
performed in the canonical ensemble 
\begin{equation}
p(E|\beta)=\frac{\rho(E)e^{-\beta E}}{Z(\beta)},
\end{equation}
 where $\rho(E)$ is the density of states and $Z(\beta)$ is the
usual normalization constant for a given $\beta$. While in the so-called
type-A superstatistics formulation $\tilde{p}(E|\beta)=\rho(E)e^{-\beta E}$
and the normalization of $p(E)$ in the Eq.(\ref{pe_init}) is performed
\textit{a posteriori}, i.e. $p(E)=\tilde{p}(E)/Z$, where $Z$ is
given by 
\begin{equation}
Z=\intop_{E}\intop\tilde{p}(E|\beta)p(\beta)d\beta\, dE.
\end{equation}
 Then, every time that an \textit{ansatz} may be assumed from the
beginning for $p(E|\beta)$ the type-B formulation is considered as
more convenient. Different functional forms of $f(\beta)$ have been
presented and succeeded in describing many complex physical systems.
Among them, we find applications as diverse as in hydrodynamic turbulent
flows \cite{beck-98}, traffic delays on the British railway network
\cite{briggs}, survival-time of cancer patients \cite{chen-08},
stock market returns \cite{kozaki-08} and quark-gluon plasma phenomenology
(see \cite{Frigori-2014}, and references therein).

The choice of $f(\beta)$ we will exploit in this study is known as
the $\chi^{2}$-distribution 
\begin{equation}
f(\beta)=\frac{1}{\beta_{0}}\frac{c^{c}}{\Gamma(c)}\left(\frac{\beta}{\beta_{0}}\right)^{c-1}\,{\rm exp}\left(-\frac{c\beta}{\beta_{0}}\right),\label{chi2_distribution}
\end{equation}
 where constants $\beta_{0}>0$ and, $n=2c$ is the number of degrees
of freedom of the system. In particular, it deserves to be noted that
Eq.(\ref{chi2_distribution}) may partially recover the so-called
Tsallis nonextensive statistics when the constant $c$ is formally
related to the Tsallis parameter $q$ by $c=1/(q-1)$. The constant
$\beta_{0}$ is related to the average and variance of the spatio-temporal
fluctuations of physical quantity $\beta$, once $\langle\beta\rangle_{f}=\intop_{0}^{\infty}\beta f(\beta)\, d\beta=\beta_{0}$
and ${\rm Var}(\beta)=\langle\beta^{2}\rangle_{f}-\beta_{0}^{2}=\beta_{0}^{2}/c$.
Still, any coupling of a physical system to finite thermal baths would
be expected to be properly described by this somehow interpolating
framework, given that $\beta_{0}$ can even be identified with a sharply
defined inverse temperature in limit when $Var(\beta)\rightarrow0$
i.e., if coupled to a thermal reservoir.

This last property would be specially desirable to better understand
the thermal behavior of a large set of systems endowed by long-range
interactions \cite{barre-01,mukamel-05,barre-05,campa-07} and, whose
microcanonical and canonical thermodynamical properties present notable
inequivalence. It is broadly accepted that those aspects arise as
consequences of the nonconcavity of the entropy, seen as a function
of the energy \cite{barre-01,bouchet-05,casetti-07}. Thus, within
an interval $(\varepsilon_{a},\varepsilon_{b})$ of energies where
the microcanonical entropy is not a concave function, the microcanonical
and the canonical ensembles become nonequivalent. A word of caution
may be due here, once we distinguish between the microcanonical entropy
$S_{\mu}\left(\varepsilon\right)$ and the canonical one $S_{can}\left(\beta\right)$,
as the latter is obtained as the Legendre-Fenchel transform of the
free energy $\varphi(\beta)$, an operation that always yields a concave
function of $\beta.$

To circumvent that technical hindrance, Touchette \textit{et al.}
have recently presented (\cite{Touchette-2010}, and references therein)
a series of methods to analytically calculate entropies that are nonconcave
functions of the energy. This can be implemented by some generalized
canonical ensembles \cite{Costeniuc-2005}, as the Gaussian Ensemble
\cite{Hetherington1,Hetherington2} or its extended version (EGE)
\cite{Johal}, where the variance of temperature is parameterized
$\left(\gamma\right)$ proportionally to the inverse thermal capacity
of the heat bath. Therefore, by taking proper limits in this unified
approach one can recover usual results in different ensembles \cite{BEC_EGE},
even when they are inequivalent in the thermodynamic limit \cite{barre-01}.

Thereby, maybe it comes as a startling remark upon aforesaid generalized
ensemble approaches, as the $\chi^{2}$-superstatistics and EGE, that
their universal thermodynamic equivalence (in the sense of Costeniuc
\cite{Costeniuc-2005}) may not be ensured right from the beginning.
Then, the quest for the existence of equivalent thermodynamic descriptions
of some peculiar physical systems studied under different generalized
ensembles can only be set by examining their explicit solutions. This
is the purpose of our present article, where a $\chi^{2}$-superstatistics
is employed to investigate by explicit calculations the infinite-range
Blume-Capel (BEC) model, which is a well-known representative system
displaying inequivalence of canonical and microcanonical phase diagrams
and, whose EGE solution was lately provided \cite{BEC_EGE}.

To this end, the article is organized as follows. A short theoretical
review on generalized canonical ensembles is provided in Section 2,
where EGE is highlighted as the most eminent representative in a class
of interpolating canonical $\leftrightarrow$ microcanonical ensembles.
Section 3 introduces the foundations of a particular limit of superstatistics,
whose choice of $f(\beta)$ is known as $\chi^{2}$- distribution.
Here it is emphasized that, most of times, this formulation is equivalent
and so allows for recovering physical results derived from the nonextensive
Statistical Mechanics of Tsallis \cite{Four_versions_Tsallis,Tsallis}.
In the Section 4 explicit analytic solutions of the BEC model is provided
in the context of $\chi^{2}$-superstatistics. Finally, Section 5
is dedicated to our concluding remarks. For completeness, a sketch
of solution for the BEC model is worked out in the canonical ensemble
in Appendix A, once its partition function $Z_{can}$ is widely employed
along this article.

\section{Generalized canonical ensembles}

The idea underlying the foundations of generalized ensembles is to
enable the computation of the microcanonical entropy $S_{\mu}\left(E\right)$
\cite{Gross_book} from a Legendre transform of a generalized free
energy function (for instance, see \cite{Touchette-2010,Costeniuc-2005})
\begin{equation}
\varphi_{g}\left(\beta\right)=-\lim_{N\rightarrow\infty}\frac{1}{N}\ln Z_{N,g}\left(\beta\right),\label{Phi_g}
\end{equation}
 whereas use is made of a generalized partition function 
\begin{equation}
Z_{N,g}\left(\beta\right)=\int e^{-\beta H_{N}\left(\sigma\right)-Ng\left(H_{N}\left(\sigma\right)/N\right)}d\sigma,\label{Z_N_gamma}
\end{equation}
formulated with the help of $g$ as a function of the mean energy
per degree of freedom $H_{N}\left(\sigma\right)/N.$ Then, if a proper
choice of $g$ can be made so implying that $\varphi_{g}\left(\beta\right)$
is differentiable at $\beta,$ the microcanonical entropy shall be
recovered by taking a generalized Legendre transform 
\begin{equation}
S_{\mu}\left(E\right)=\beta E-\varphi_{g}\left(\beta\right)+g\left(E\right),\label{Generalized_Legendre}
\end{equation}
 where the constraint $E=\varphi'_{g}\left(\beta\right)$ shall be
fulfilled.

In particular, when taking $g\left(E\right)=\tilde{\gamma}\, E^{2}/2$
one is straightforwardly led to the simplest form of the so-called
Extended Gaussian Ensemble \cite{Johal}, whose partition function
$Z_{N,\tilde{\gamma}}\left(\beta\right)$ may be derived from Eq.(\ref{Z_N_gamma})
with the help of some usual Gaussian integrals as 
\begin{equation}
e^{-\tilde{\gamma}E^{2}/2}=\sqrt{\frac{\tilde{\gamma}}{2\pi}}\intop_{-\infty}^{\infty}e^{-\tilde{\gamma}x^{2}/2-i\tilde{\gamma}Ex}dx.\label{Gaussian_integral}
\end{equation}
 Therefore, one can interpret $Z_{N,\tilde{\gamma}}\left(\beta\right)$
as an integral transform \cite{Touchette-2010} of the canonical
partition function, 
\begin{equation}
Z_{N,\tilde{\gamma}}\left(\beta\right)=\sqrt{\frac{\tilde{\gamma}N}{2\pi}}\intop_{-\infty}^{\infty}e^{-\tilde{\gamma}Nx^{2}/2}Z_{can}\left(\beta+i\tilde{\gamma}x\right)dx.\label{Z_EGE}
\end{equation}

Unfortunately, those integrals are generally not ease to compute for
physically sound systems. So, explicit calculations of this kind as
performed for the infinite-range BEC model \cite{BEC_EGE}, are recognized
as quite scarce. In addition, it is illustrative to note that the
usual canonical partition function $Z_{can}\left(\beta\right)$ may
be trivially recovered by EGE in the $\tilde{\gamma}\rightarrow0$
limit, i.e. $Z_{can}\left(\beta\right)=\lim_{\tilde{\gamma}\rightarrow0}Z_{N,\tilde{\gamma}}\left(\beta\right).$
While due to delta sequence representation $\delta_{\tilde{\gamma}}\left(E\right)=\sqrt{\tilde{\gamma}/\pi}e^{-\tilde{\gamma}E^{2}}$one
directly recovers $S_{\mu}\left(E\right)$ when $\tilde{\gamma}\rightarrow\infty$
from aforesaid standard methods.

\section{Superstatistics: the $\chi^{2}$-distribution}

Let us define $U$ as the mean energy of the entire superstatistical
system consisting of many cells. Since we are assuming that each cell
has an approximately constant inverse temperature $\beta,$ the energy
distribution follows from the usual Boltzmann factor $e^{-\beta E(\sigma)},$
where $E(\sigma)$ is the energy of a given state $\{\sigma\}$ in
the cell. The marginal energy distribution for a system with density
of states $\rho(E)$ becomes 
\begin{equation}
p(E;U)\sim\intop_{0}^{\infty}\,\rho(E)e^{-\beta(E-U)}f(\beta)d\beta.
\end{equation}
Here, we follow the type-A prescription to obtain the normalization
constant, which is now dependent on the constant $U$. The particular
choice for the probability density $f(\beta)$ as the $\chi^{2}$-distribution
for the fluctuating $\beta$ with mean value $\beta_{0}$ and parameter
$c$ yields the superstatistical version of the Boltzmann-Gibbs (BG)
statistics for the physical system. Defining $\theta=\beta/\beta_{0}$,
one obtains 
\begin{equation}
p(E;U)\sim\intop_{0}^{\infty}\,\rho(E)w_{c}(\theta)\, e^{-\theta\beta_{0}(E-U)}d\theta,\label{pe_Beck}
\end{equation}
 where the weight function $w_{c}(\theta)$ is given by 
\begin{equation}
w_{c}(\theta)=\frac{c^{c}}{\Gamma(c)}\theta^{c-1}e^{-c\theta}.
\end{equation}
 Then, recalling the following mathematical result 
\begin{equation}
\int_{0}^{\infty}w_{c}(\theta)\, e^{-\theta\beta E}d\theta=\left(1+\frac{\beta E}{c}\right)^{-c},
\end{equation}
 valid for $c\ge1$ and $c+\beta E>0$, the stationary energy distribution
becomes 
\begin{equation}
p(E;U)=\frac{1}{Z_{c}(\beta_{0},U)}\,\rho(E)\,\left[1+\frac{\beta_{0}\left(E-U\right)}{c}\right]_{+}^{-c},\label{pe_norm_Beck}
\end{equation}
 where $[x]_{+}={\rm max}(x,0)$ and $Z_{c}(\beta_{0},U)$ is the
partition function version of BC $\chi^{2}$-superstatistics, 
\begin{equation}
Z_{c}(\beta_{0},U)=\intop_{E\in I}\rho(E)\,\left[1+\frac{\beta_{0}\left(E-U\right)}{c}\right]_{+}^{-c}dE,
\end{equation}
 where $I$ is the energy range that may be limited by some energy
cutoff for confined systems.

At this stage it is tempting to identifying $c$ with the Tsallis
parameter $q$ by $c=1/(q-1)$, which limits this parameter to $1\le q\le2$
in type-A prescription. The value $q=2$ is allowed for systems with
finite energy range. Thus, the partition function version of the $\chi^{2}$-superstatistics
in terms of the parameter $q$ becomes 
\begin{equation}
Z_{q}(\beta_{0},U)=\intop_{E\in I}\rho(E)\, e_{q}^{-\beta_{0}(E-U)}\, dE,\label{Z_Beck}
\end{equation}
 where the notation $e_{q}^{-x}=\left[1+(q-1)x\right]^{1/(1-q)}$
for the $q$-exponential function is introduced, but needs to be restricted
to the above domain in $q$.

Such description is analogous to the one where ``escort probabilities''
are not employed \cite{Four_versions_Tsallis}. From a formal point
of view, the canonical limit is easily recovered in the $q\rightarrow1$
limit: $e_{q}(x)\rightarrow{\rm exp}(x)$. This limit describes a
system with a very large number of independent degrees of freedom
$n$ and turns the $\chi^{2}$-distribution into a Dirac delta function
$\delta(\beta-\beta_{0})$ where $\beta_{0}$ has to be defined by
a large heat bath which provides the constant temperature. Such prior
PDF contrasts, for example, to an uniform prior PDF when little is
known about the temperature of the system.

\subsection{An analytical continuation in $\beta$}

Next, we demonstrate how to obtain the complementary values for $q$
$(q<1,q>2)$ in superstatistics. However, in this case, one needs
to explore the analytical continuation for the inverse of the temperature
distribution $\beta$: $f(\beta)\rightarrow f(z)$. Thus, we start
with the contour integral representation for the reciprocal gamma
function due to Hankel to obtain the complex normalized version of
$f(\beta)$, 
\begin{equation}
f(z)=\frac{i}{2\pi\beta_{0}}\frac{\Gamma(d)}{c^{d-1}}\left(\frac{-z}{\beta_{0}}\right)^{-d}\,{\rm exp}\left(-\frac{c\, z}{\beta_{0}}\right),\label{comp-f}
\end{equation}
 with $d\ne0,-1,-2,\cdots$. The integration in $z$ needs to be performed
along a contour $\Gamma$ starting at $+\infty$ on the real axis,
going counterclockwise around the origin and back to $+\infty$ \cite{abramo}.
This derivation is similar to real $\beta$ case when one starts from
the normalized gamma function: $\frac{1}{\Gamma(c)}\intop_{0}^{\infty}t^{c-1}e^{-t}dt=1$.
Therefore, the corresponding real partition function becomes 
\begin{equation}
Z_{d}(\beta_{0},U)=\intop_{E\in I}\intop_{\Gamma}\rho(E)\,\omega_{d}(\theta)e^{-\theta\beta_{0}(E-U)}d\theta dE,
\end{equation}
 where $\theta=z/\beta_{0}$ is the new complex integration variable
and 
\begin{equation}
\omega_{d}(\theta)=\frac{i}{2\pi}\frac{\Gamma(d)}{c^{d-1}}(-\theta)^{-d}e^{-c\theta}.
\end{equation}
 The integration over the contour $\Gamma$ is easily performed and
one obtains 
\begin{equation}
Z_{d}(\beta_{0},U)=\intop_{E\in I}\rho(E)\,\left[1+\frac{\beta_{0}\left(E-U\right)}{c}\right]_{+}^{d-1}\, dE.
\end{equation}
 The identifications $c=1/(q-1)$ and $d-1=1/(1-q)$ yield the known
results of the $q$-statistics but now the parameter $q$ is restricted
to $q<1$ and $q>2$. This formal construction yields the complementary
range of validity in the values of $q$ for the $\chi^{2}$-superstatistics.
It is quite similar to the integral parameterization of the Tsallis
partition function $Z_{q}$ used by Prato \cite{prato} to extend
the so-called Hilhorst formula ($q>1$) to $q<1$.

\section{Beck-Cohen $\chi^{2}$-superstatistics solutions of BEC model}

The superstatistics formulation for nonequilibrium systems described
by the fluctuation parameter $\beta$ according to the $\chi^{2}$-distribution
has energy distribution given by Eq. (\ref{pe_Beck}), which can be
cast in the form of the following integral transform 
\begin{equation}
p(E;U)\sim\intop_{0}^{\infty}\, w_{c}(\theta)\, e^{\theta\beta_{0}U}Z_{can}(\theta\beta_{0})d\theta.
\end{equation}
 The BEC model is described by $Z_{can}(\beta)$ given in Eq. (\ref{App-A3})
for a system with $N$ spins and average energy $U$. The thermodynamic
free energy density version of BC superstatistics is defined as 
\begin{equation}
\varphi_{c}(\beta_{0},\varepsilon)=-\lim_{N\rightarrow\infty}\frac{1}{N}\ln Z_{c}(\beta_{0},\varepsilon),
\end{equation}
 where $Z_{c}(\beta_{0},\varepsilon)$ is the normalization factor
in Eq. (\ref{Z_Beck}) written as a function of the average energy
per spin $\varepsilon=U/N\Delta$. 

As usual, the partition function $Z_{c}(\beta_{0},\varepsilon)$ can
be evaluated in the large $N$ limit by means of the saddle-point
approximation, as shown in Eq. (\ref{App-A4}) for the canonical ensemble.
Afterwards, the integration over $\theta$ is analytically performed,
which introduces an extra factor that depends on $\varepsilon$ and
$c$: $(1-\beta_{0}\varepsilon N\Delta/c)^{c}$. Thus, in terms of
$c=1/(q-1)$ and a newly introduced constant $\gamma=N(q-1)=N/c$,
the free energy density is given by 
\begin{figure}[t]
\begin{centering}
\includegraphics[width=8.5cm]{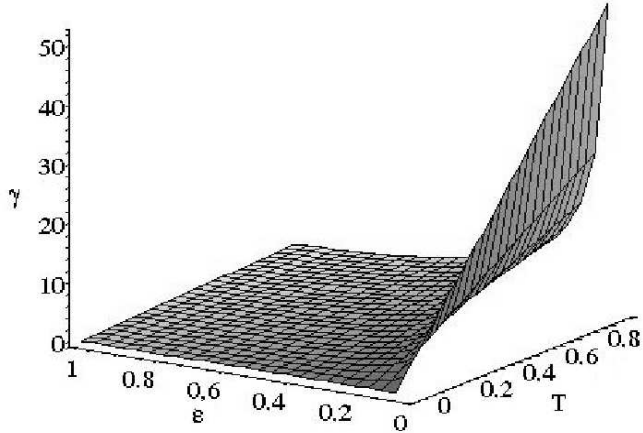} \caption{Maximum values allowed for $\gamma$ according to Eq. (\ref{eq:vinculosB})
as a function of Energy $\left(\varepsilon\right)$ and Temperature
$\left(T\right)$ when $\Delta/J=0.462407$.}

\par\end{centering}

\label{fig:1} 
\end{figure}
\begin{figure}
\includegraphics[width=7.2cm]{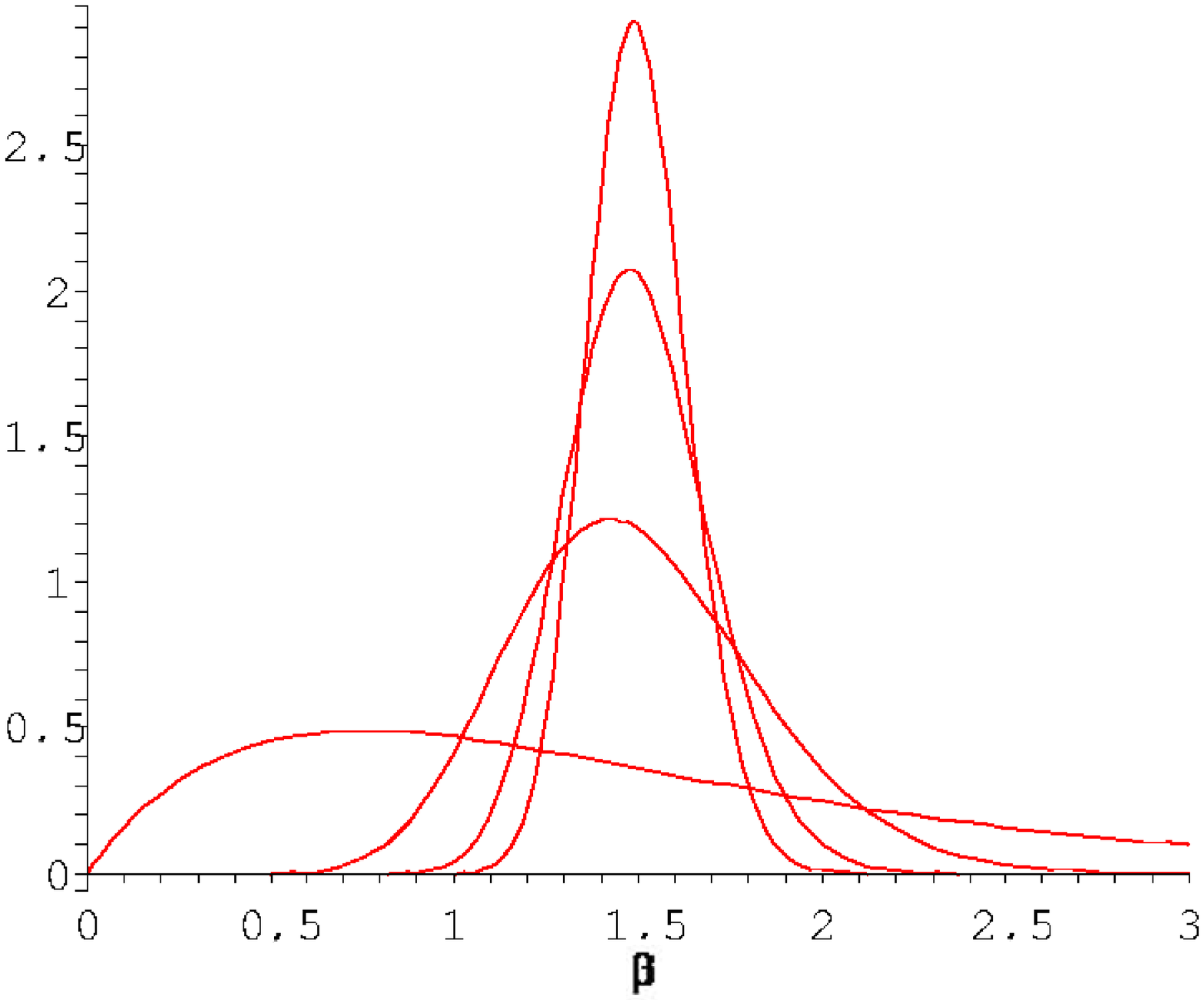}\includegraphics[width=7.2cm]{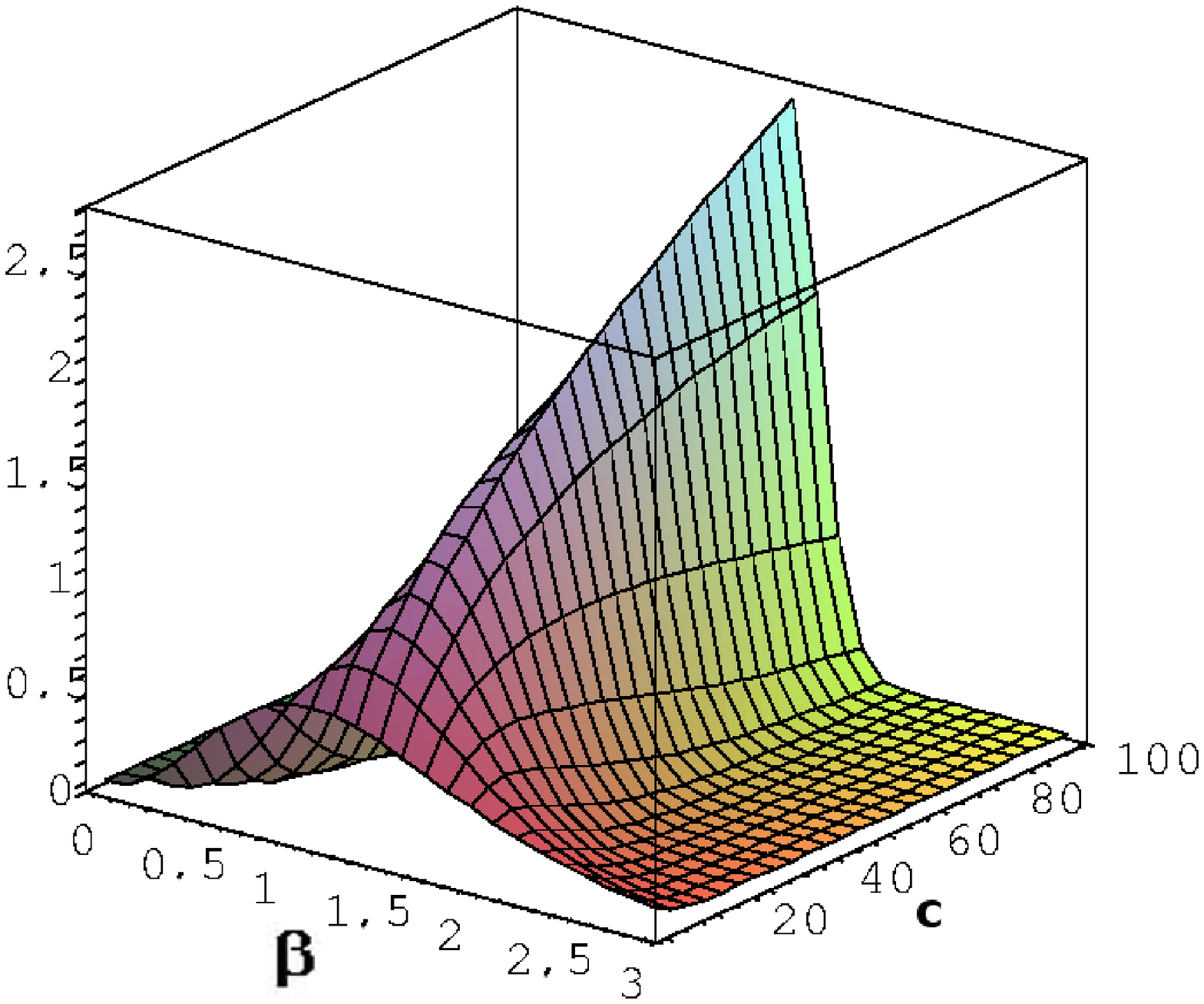}
\caption{LEFT PANEL: The $\chi^{2}$-distribution as given by Eq.(\ref{chi2_distribution})
as a function of $\beta$ and increasing values of $c=\left\{ 2,20,60,120\right\} $
for fixed $\beta_{0}=3/2.$ RIGHT PANEL: the same distribution but
in a 3d perspective. Note that when $c\rightarrow\infty$ a Dirac
delta distribution is recovered.}
\end{figure}
 
\begin{equation}
\begin{array}{cc}
\varphi_{\gamma}^{*}(\beta_{0},\varepsilon,m,p)= & \frac{1}{\gamma}\ln\left[1+\gamma\beta_{0}\Delta\left(p-Km^{2}-\varepsilon\right)\right]-\frac{1}{\gamma}\ln\left(1-\gamma\beta_{0}\Delta\varepsilon\right)\\
 & +\ln\left\{ \left[\frac{\sqrt{p^{2}-m^{2}}}{2\left(1-p\right)}\right]^{p}\left[\frac{p+m}{p-m}\right]^{m/2}(1-p)\right\} .
\end{array}\label{Free_energy_chi2_BEC}
\end{equation}
The notation $\varphi^{*}$ means that $\varphi$ is evaluated at
its stationary points $(m^{*},p^{*}),$ whereas the variables $\varepsilon,\, m$
and $p$ are considered to be independent. While the third term of
Eq.(\ref{Free_energy_chi2_BEC}) is straightforwardly expressed as
$S_{\mu}\left(\varepsilon\right)/N$ \cite{mukamel-05,BEC_EGE},
the microcanonical constraint $\varepsilon=p-Km^{2}$ is never enforced
in this approach. Still, it is worthy to note that the integration
over $\theta$ leads to the constraints 
\begin{eqnarray}
-1 & < & \gamma\beta_{0}\Delta(p-Km^{2}-\varepsilon)\label{eq:vinculosA}\\
\gamma\beta_{0}\Delta\varepsilon & < & 1,\label{eq:vinculosB}
\end{eqnarray}
 which must be satisfied together with other saddle point conditions.

Figure 1 shows the maximum values of $\gamma$ that satisfies the
condition (\ref{eq:vinculosB}) as a function of $\varepsilon$ and
$T_{0}=1/\beta_{0}$. In particular, once the BG statistics exhibits
a first-order phase transition at $(T\simeq0.330666,\varepsilon\simeq0.330,$
see \cite{mukamel-05,BEC_EGE}) for the coupling $\Delta/J=0.462407$,
one shall not exceed $\gamma\sim2.17$ if it is expected to recover
the BG thermodynamic results in the large $N$ limit, given that $f(\beta)\rightarrow\delta(\beta-\beta_{0})$
as a consequence of $Var(\beta)\rightarrow0$ when $N\rightarrow\infty$.
This limiting behavior is followed by the mode of the distribution
tending to $\langle\beta\rangle_{f}$ because the maximum of $f(\beta)$
occurs at $\beta=\beta_{0}(1-2/N)$, as seen in Figure 2.

By keeping $\gamma$ constant $\varphi^{*}$ is independent of $N$
and in this case as the number of spins increases $q$ goes asymptotically
to 1. Accordingly, for a finite and small $\gamma$ one is allowed
to Taylor expand Eq. (\ref{Free_energy_chi2_BEC}) as a functions
of that parameter, 
\begin{equation}
\begin{array}{cc}
\varphi_{\gamma}^{*}(\beta_{0},\varepsilon,m,p) & =\beta_{0}\Delta\left(p-Km^{2}\right)+\frac{\gamma}{2}\beta_{0}^{2}\,\Delta^{2}\left[\varepsilon^{2}-\left(p-Km^{2}-\varepsilon\right)^{2}\right]\\
 & +\frac{\gamma^{2}}{3}\beta_{0}^{3}\,\Delta^{3}\left[\varepsilon^{3}+\left(p-Km^{2}-\varepsilon\right)^{3}\right]+\mathsf{\mathcal{O}}(\gamma^{3})\\
 & +\ln\left\{ \left[\frac{\sqrt{p^{2}-m^{2}}}{2\left(1-p\right)}\right]^{p}\left[\frac{p+m}{p-m}\right]^{m/2}(1-p)\right\} .
\end{array}\label{Taylor_Free_energy_chi2}
\end{equation}
Surprisingly, up to its lowest orders the expansion in Eq. (\ref{Taylor_Free_energy_chi2})
closely resembles the exact solution formerly obtained in the EGE
framework \cite{BEC_EGE}, where the free energy was given by 
\begin{equation}
\begin{array}{cc}
\varphi_{\tilde{\gamma}}(\beta_{0},\varepsilon,m,p) & =\beta_{0}\Delta(p-Km^{2})+\tilde{\gamma}\Delta^{2}(p-Km^{2}-\varepsilon)^{2}+\\
 & \ln\left\{ \left[\frac{\sqrt{p^{2}-m^{2}}}{2\left(1-p\right)}\right]^{p}\left[\frac{p+m}{p-m}\right]^{m/2}(1-p)\right\} .
\end{array}\label{Phi_EGE_BEC}
\end{equation}
There $\tilde{\gamma}$ denotes the EGE free-parameter once explicit
in Eq.(\ref{Z_EGE}) and, by whose tuning it was shown that the extremum
of $\varphi_{\tilde{\gamma}}(\beta_{0},\varepsilon,m,p)$ was able
to interpolate between solutions in the canonical ($\tilde{\gamma}=0$)
and microcanonical ensembles ($\tilde{\gamma}\rightarrow\infty$).
Thus, by studying the analytic solutions dependent on $\gamma$ (and
$\tilde{\gamma}$) we may infer that $\chi^{2}$-superstatistics and
EGE approaches would perhaps produce equivalent thermodynamic descriptions
for the BEC model%
\footnote{It is in principle expected that for ``small'' values of $\gamma$
and $\tilde{\gamma}$ the Eq. (\ref{Taylor_Free_energy_chi2}) and
(\ref{Phi_EGE_BEC}) becomes equivalent till second order. But, in
fact, it comes as a surprise that the thermodynamics emerging from
both solutions is the same even when $\tilde{\gamma}\rightarrow\infty$
but while $\gamma\sim0.$%
}. This hypothesis was earlier proposed in a broader and more abstract
perspective by Johal \cite{Johal} and Morishita \cite{Morishita-2006},
but till now it has been waiting to be corroborated by an explicit
calculation.

\begin{figure}[!ht]
\centering{}\includegraphics[width=12cm]{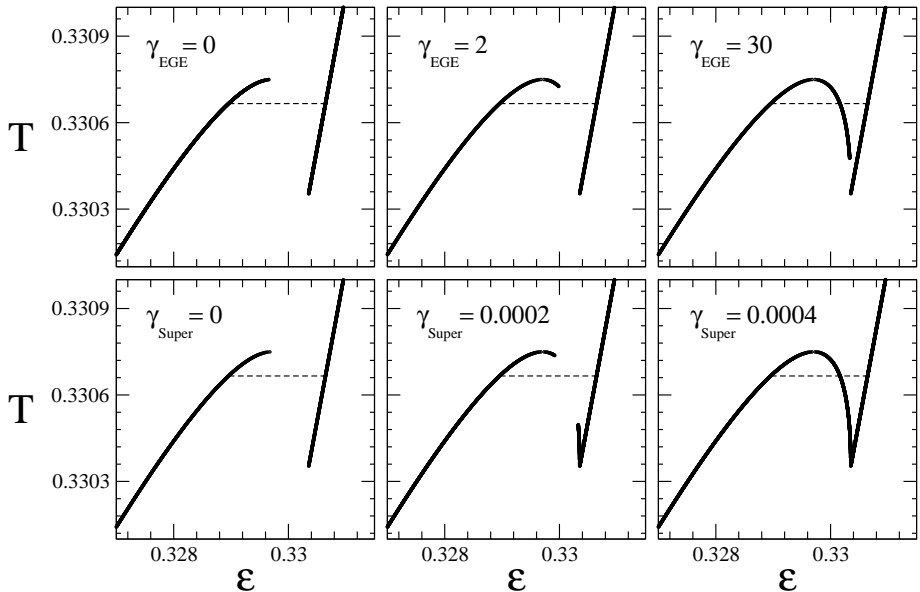} \caption{Caloric curves $T(\varepsilon)\times\varepsilon$ emerging from the
minimization of EGE (upper) and $\chi^{2}$-superstatistics (lower)
solutions, depicted respectively from Eqs. (\ref{Phi_EGE_BEC}) and
(\ref{Free_energy_chi2_BEC}). The BEC coupling is set to $\Delta/J=0.462407,$
which implies on a first order phase transition in the canonical ensemble
with a critical temperature $T\simeq0.330666$, seen as dashed lines
(Maxwell construction). By increasing values of parameters for EGE
ensemble $\left(\gamma_{EGE}\right)$ and $\chi^{2}$-superstatistics
$\left(\gamma_{Super}\right)$ the microcanonical caloric curve is
gradually recovered, so corroborating the thermodynamic equivalence
of both ensembles.}
\end{figure}

Therefore, to investigate the full thermodynamics of BEC model in
our BC approach, we have numerically studied the behaviour of $\varphi_{\gamma}^{*}(\beta_{0},\varepsilon,m,p)$
at the points $m^{*}$ and $p^{*}$ that minimize $\varphi$ as a
function of parameters $\beta_{0}$ and $\varepsilon$ for a set of
$\gamma$ values. We observe that the limiting case $\gamma\rightarrow0$
trivially yields the canonical ensemble results (\ref{App-A5}). Additionally,
it was verified how large the parameter $\gamma$ might be set, while
obeying the constraints from Eqs. (\ref{eq:vinculosA}) and (\ref{eq:vinculosB}),
to possibly recover the well-known microcanonical limit. A summary
of our studies is depicted in Figure 3, where the caloric curves $T\left(\varepsilon\right)\times\varepsilon$
extracted from EGE and $\chi^{2}$-superstatistics solutions are plotted
for various values of $\gamma$ (or, respectively $\tilde{\gamma}$)
for comparative purposes. We observe that in contradistinction to
EGE paradigm where microcanonical results would be recovered as a
limiting case $\tilde{\gamma}\rightarrow\infty,$ the $\chi^{2}$-superstatistics
had not to rely on analogous formal limits and, in fact, it was able
to fully generate the microcanonical regime for quite small values
of $\gamma.$ So, we verify that $\chi^{2}$-superstatistics and EGE
can be considered as successful interpolating approaches for Statistical
Mechanics which present ensemble equivalence at the thermodynamic
level.

\section{Concluding remarks}

We have evaluated the free energy for the infinite-range BEC model,
a system of spins that does not present equivalence between usual
(i.e. canonical and microcanonical) ensembles of Statistical Mechanics
by employing two generalized statistical frameworks, to know, the
$\chi^{2}$-superstatistics and EGE. 

Both generalized ensembles here employed were primarily devised to
describe systems in contact with finite thermal baths, so presenting
peculiar kinds of stationary thermal behavior. Then, thermal fluctuations
on those heat baths can be addressed within the superstatistics approach
by assuming a probability distribution function $f(\beta)$, which
we take as being a $\chi^{2}$-distribution, inspired by the fact
that this distribution reproduces thermodynamic results resembling
the well-known Tsallis nonextensive statistics for various regimes
of $q$. On the other hand, the EGE approach is based on the assumption
that gaussian thermal fluctuations happen in the vicinity of thermal
equilibrium, once it is established by physical systems coupled to
finite reservoirs. This was proved to consists on a powerful tool
for studying systems with nonconcave entropies as proposed in Ref.
\cite{Touchette-2010}. 

Actually, it would be reasonable to expect that depending on the physical
system studied by both aforesaid statistical frameworks the resulting
thermodynamic predictions might even differ, at least in specific
regimes \cite{Morishita-2006}, as supposedly seen from some experiments
on hadronic production processes\cite{Osada_2006}. However, it was
amazing to find that both our exact solutions for the infinite-range
BEC model, obtained in the EGE and $\chi^{2}$-superstatistics, were
able to provide us with the same thermodynamical outcome in all conceivable
physical regimes described by tuning $\gamma$ or $\tilde{\gamma}$.
That is, when $\gamma=\tilde{\gamma}=0$ the usual canonical solution
was restored as a limiting case by both generalized ensembles, while
by taking $\tilde{\gamma}\rightarrow\infty$ the EGE solution converged
to the microcanonical case: a regime that $\chi^{2}$-superstatistics
was also able to reach, but already at very small values of $\gamma.$
\\

\textbf{Acknowledgements}

The authors acknowledge support by FAPESP and CAPES (Brazil).

\appendix

\section{The Blume-Capel model in the canonical ensemble}

Let us consider the mean-field version of the BEC model for $N$ spins
$S_{i}$, 
\begin{equation}
E(S)=\Delta\sum_{i=1}^{N}S_{i}^{2}-\frac{J}{2N}\left(\sum_{i=1}^{N}S_{i}\right)^{2}\,,\label{App-A1}
\end{equation}
 where $S_{i}=\{0,\pm1\}$. The couplings $J>0$ and $\Delta$ are
the exchange and crystal-field interactions, respectively. The BEC
model represents a simple generalization of the spin-$1/2$ Ising
model, but with a rich phase diagram in the $(\Delta/J,T/J)$ plane.
It exhibits a first-order transition line, tricritical point, and
a second-order transition line.

It is useful to introduce the order parameters magnetization $M=\sum_{i=1}^{N}S_{i}=N_{+}-N_{-}$
and its second moment, the quadrupole moment $P=\sum_{i=1}^{N}S_{i}^{2}=N_{+}+N_{-}$,
where $N_{+}$ and $N_{-}$ are, respectively, the number of sites
with up and down spins to describe this model. If $N_{0}$ is defined
as the total number of zero spins, then $N=N_{+}+N_{-}+N_{0}$ is
the total number of spins in the system, then Eq. (\ref{App-A1})
is rewritten as 
\begin{equation}
E=\Delta P-\frac{J}{2N}M^{2}.\label{App-A2}
\end{equation}

After summing up over all configuration states%
\footnote{It is convenient to obtain the canonical partition function $Z_{can}$
of this model by taking the limit $\gamma\rightarrow0$ in \cite{BEC_EGE},
where $\gamma$ can be considered just as a parameter to perform an
integral regularization procedure \cite{Lukkarinen_99}.%
}, we arrive to the following formal canonical partition function described
in terms of the conditioned order parameters $P$ and $M$ instead
of the number of spins $N,N_{+}$ and $N_{-}$, 
\begin{eqnarray}
Z_{can}(\beta) & = & \sum_{P=0}^{N}\sum_{M=-P}^{P}\frac{N!}{\left(N-P\right)!\left[\frac{1}{2}\left(P-M\right)\right]!\left[\frac{1}{2}\left(P+M\right)\right]!}\, e^{-\beta E\left(M,P,N\right)}\\
 & \equiv & \sum_{P=0}^{N}\sum_{M=-P}^{P}\Omega(M,P,N)\, e^{-\beta E\left(M,P,N\right)}.\label{App-A3}
\end{eqnarray}
 As usual, $\beta$ stands for the inverse of thermodynamic temperature
and we take the Boltzmann contant $k_{B}=1$.

To evaluate the above expression for large $N$ limit, we consider
the variational approach (saddle-point approximation), which turns
the double sum over configurations of $Z_{can}$ into 
\begin{equation}
Z_{can}(\beta)\simeq e^{-N\varphi^{*}(\beta,m,p)},\label{App-A4}
\end{equation}
 where $\varphi^{*}$ means that the free energy is asymptotically
evaluated for large $N$ around the stationary points $m^{*}$ and
$p^{*}$. For such calculation, it is convenient to work with intensive
quantities $m=M/N$ and $p=P/N$, and define $K=J/(2\Delta)$. Therefore,
performing the expansion around the stationary point up to second
order \cite{Gross_book} and using the Stirling's approximation $\ln N!\simeq N\ln N-N$,
one obtains the free energy, 
\begin{equation}
\varphi^{*}(\beta,m,p)=\beta\Delta(p-Km^{2})+\ln\left\{ \left[\frac{\sqrt{p^{2}-m^{2}}}{2\left(1-p\right)}\right]^{p}\left[\frac{p+m}{p-m}\right]^{m/2}(1-p)\right\} ,\label{App-A5}
\end{equation}
 where the notation $\varphi^{*}$ also implies that the points $m$
and $p$ are solutions of the minimization equations 
\begin{equation}
\frac{\partial\varphi}{\partial m}=\frac{\partial\varphi}{\partial p}=0,\label{App-A6}
\end{equation}
 as well as must satisfy the stability condition 
\begin{equation}
\det\left(\begin{array}{cc}
\frac{\partial^{2}\varphi}{\partial m^{2}} & \frac{\partial^{2}\varphi}{\partial m\partial p}\\
\frac{\partial^{2}\varphi}{\partial p\partial m} & \frac{\partial^{2}\varphi}{\partial p^{2}}
\end{array}\right)\geq0,\label{App-A7}
\end{equation}
 for fixed energies $\varepsilon=E/\Delta N$. Equations (\ref{App-A6})
and (\ref{App-A7}) is all one needs to describe the thermodynamic
behavior of the BEC model. We also emphasizes that $p$ and $m$ are
independent variables. Thus, the constraint $\varepsilon=p-Km^{2}$
is not enforced, contrary to the thermodynamic description in the
microcanonical approach. \\

\end{document}